# Real-time computation of bio-heat transfer in the fast explicit dynamics finite element algorithm (FED-FEM) framework

Jinao Zhang, Sunita Chauhan

Department of Mechanical and Aerospace Engineering, Monash University, Wellington Rd, Clayton, VIC 3800 Australia

**Abstract.** Real-time analysis of bio-heat transfer is very beneficial in improving clinical outcomes of hyperthermia and thermal ablative treatments but challenging to achieve due to large computational costs. This paper presents a fast numerical algorithm well suited for real-time solutions of bio-heat transfer, and it achieves real-time computation via the (i) computationally efficient explicit dynamics in the temporal domain, (ii) element-level thermal load computation, (iii) computationally efficient finite elements, (iv) explicit formulation for unknown nodal temperature, and (v) pre-computation of constant simulation matrices and parameters, all of which lead to a significant reduction in computation time for fast run-time computation. The proposed methodology considers temperature-dependent thermal properties for nonlinear characteristics of bio-heat transfer in soft tissue. Utilising a parallel execution, the proposed method achieves computation time reduction of 107.71 and 274.57 times compared to those of with and without parallelisation of the commercial finite element codes if temperature-dependent thermal properties are considered, and 303.07 and 772.58 times if temperature-independent thermal properties are considered, far exceeding the computational performance of the commercial finite element codes, presenting great potential in real-time predictive analysis of tissue temperature for planning, optimisation and evaluation of thermo-therapeutic treatments. The source code is available at https://github.com/jinaojakezhang/FEDFEMBioheat.

## 1. Introduction

Fast and accurate modelling of bio-heat transfer in soft biological tissue is increasingly valuable in the applications of computer-assisted surgery, in particular, the tissue temperature prediction and thermal dose computation for therapeutic hyperthermia [1], thermal damage evaluation for tissue ablation [2, 3], and the development of associated surgical simulators for cancer treatment [4, 5]; however, it is challenging to satisfy the conflicting requirements (fast and accurate). The patient outcome of thermo-therapeutic treatment is dependent on the precise control of thermal energy to induce desired selective thermal injury to the tissue, forming a tissue coagulation zone to sufficiently cover the tumour region without affecting surrounding healthy tissue. Excessive thermal energy may lead to complications such as burns to nearby organs and/or recurrence of tumours due to insufficient thermal energy. In recent studies, it was shown that the real-time prediction and control of thermal dose could avoid thermally hot-spots in healthy tissue, leading to improvements in clinical outcomes [6, 7]. However, most of the developed numerical methods [8-13] are mainly focused on numerical accuracy, convergence, and stability, rather than computation time to satisfy the conflicting requirements for better clinical outcomes. This is mainly hindered by the computational challenge for real-time computation of thermal dose, involving large computational costs for solving the bio-heat transfer equation for tissue temperature field. For instance, it was reported that simulating a thermal ablation process using conventional methods led to a computation time ranging from a few minutes to several hours [14], and that simulation of an 8-minute thermal ablation took around 6.0 hours [15].

The challenge of a solution method to achieve real-time computation of bio-heat transfer is mainly due to the nonlinear characteristics of bio-heat transfer in soft tissue and the computational efficiency of the solution method. It was shown that soft tissue thermal properties such as tissue density, thermal conductivity, and specific heat capacity are temperature-dependent [16-18]. Furthermore, bio-heat conduction, blood perfusion, and metabolic heat generation lead to a nonlinear bio-heat transfer behaviour in soft tissue, resulting in a nonlinear system of equations that needs to be solved at each time step. The nonlinear system of equations is usually solved by an iterative solution procedure based on the Newton–Raphson method [19] through a sequence of solutions of a linearised system of equations. The linear system of equations can either be solved by (i) directly computing the inverse or the factorisation of the system matrix or (ii) iteratively solving a system of algebraic equations based on an initial estimate; however, both solution approaches lead to a significant increase in computation time.

Various techniques were reported to facilitate the computational performance of numerical methods for solutions of the bio-heat transfer equation, and most of the existing techniques were based on the finite difference method (FDM). Schwenke et al. [20] studied a Graphics Processing Unit (GPU)-accelerated FDM to achieve fast simulation of focused ultrasound treatment via a parallel execution of the solution procedure on GPU; however, FDM requires a regular computation grid to approximate spatial derivatives, but human organs/tissue are irregular shapes with curvilinear boundaries, resulting in inaccuracy for accommodating soft tissue material properties and enforcing boundary conditions. He and Liu [21] developed a parallel alternating direction explicit (ADE) scheme based on FDM to solve the bio-heat equation; Carluccio et al. [22] devised a spatial filter method based on Fast Fourier Transform (FFT) with FDM to reduce computation time; Kalantzis et al. [23] studied a GPU-accelerated FDM for fast simulation of focused ultrasound thermal ablation; Dillenseger and Esneault [24] also studied an FFT-based FDM method; Chen et al. [25] presented a GPU-accelerated microwave imaging method based on FDM to monitor temperature in thermal therapy; Johnson and Saidel [26] studied an FDM-based methodology for fast simulation of radiofrequency tumour ablation; and Niu et al. [27] employed cellular neural networks (CNN) based on FDM for efficient estimation of tissue temperature field. Despite the improved computational performance by the above methods, they all suffer from the inaccuracy in describing the thermal effects of irregular boundary conditions due to using FDM for computation grid. Indik et al. [28, 29] employed a multi-grid method in a finite volume discretisation for fast bio-heat transfer computation owing to multi-scales of grid resolutions. Mariappan et al. [30] developed a GPU-accelerated finite element methodology (FEM) for fast simulation of bio-heat transfer in thermal ablation. Compared to FDM, the problem domain of arbitrary shapes can be accommodated by FEM without difficulty, and mesh sizes can be varied by considering the numerical accuracy required and computational resources available; material properties can be incorporated into the parametric constitutive equations and can vary over finite elements, and boundary conditions can be handled directly in the global system of equations for elements/nodes under boundary conditions. However, the utilisation of GPU for standard FEM only improves the computational performance owing to hardware but does not solve the computational efficiency of FEM fundamentally [31]. Bourantas et al. [32] presented a dynamic mode decomposition method based on meshless point collocation for fast numerical solutions of bio-heat transfer, and Kowalski and Jin [33] studied a model order reduction method based on FDM for electromagnetic phased-array hyperthermia. Despite the improved numerical efficiency by the mode decomposition/order reduction methods, the accuracy and efficiency of the algorithm are dependent on the number of reduced-basis chosen [34], and there is an inevitable energy loss due to the projection from the full system space to the smaller subspace [35]. Recently, the authors developed a fast explicit dynamics finite element algorithm (FED-FEM) for fast computation of transient heat transfer problems [36]. The FED-FEM is computationally efficient and can maintain numerical convergence and accuracy, and it is particularly suited for real-time thermal applications.

In this paper, a new methodology is developed based on the framework of FED-FEM [36] for fast simulation and analysis of bio-heat transfer in soft tissue, and it is studied using parallel implementation for real-time bio-heat computation. To achieve fast run-time computation, the proposed methodology is established with the aims to (i) eliminate the need for thermal stiffness matrix inversion during temporal integration, (ii) eliminate the need for assembling global thermal matrices of entire model, (iii) efficiently obtain spatial thermal responses, (iv) eliminate the need for conventional iterative solutions of the nonlinear system of equations and allow parallel execution, and (v) conduct computation of constant matrices and parameters before run-time computation. It aims to utilise the proposed efficient algorithm coupled with computing power of hardware for fast numerical updates at run-time computation. Consequently, it achieves the above aims by (i) utilising explicit dynamics in the temporal domain, (ii) proposing an element-level thermal load computation, (iii) employing computationally efficient finite elements, (iv) devising an explicit formulation for nodal temperature, and (v) pre-computing constant matrices and simulation parameters. The proposed methodology can accommodate non/linear thermal material properties and handle nonlinear characteristics involved in bio-heat transfer problems. The computational benefits are examined, and assessments of numerical errors are presented. Simulation results demonstrate that the proposed method can achieve real-time computations while maintaining numerical accuracy, leading to an efficient and accurate numerical algorithm for conducting bio-heat transfer analysis for real-time biomedical applications.

The remainder of this paper is organised as follows: Section 2 presents the formulation of the proposed methodology in the framework of FED-FEM. Section 3 evaluates the computational performance in terms of numerical accuracy and computation time reductions, and discussions are presented in Section 4. Finally, this paper concludes in Section 5 with future improvements to the work.

## 2. FED-FEM for Pennes bio-heat transfer

### *2.1 Pennes bio-heat transfer model*

The heat transfer in soft biological tissue may be characterised by various bio-heat transfer models, among which the most well-known is the Pennes bio-heat transfer model by Harry H. Pennes [37], which mathematically describes the heat transfer process in living biological tissue composed of solid tissue and blood flow [38]. The Pennes bio-heat transfer model can provide suitable temperature predictions in the whole body, organ, and tumour analyses [39], and therefore it has been widely used in the modelling of thermal ablation for cancer treatment [16, 23, 30, 40] and other biomedical research areas [12, 21, 41, 42].

Based on the heat flow equilibrium due to the conservation law of energy [43], the Pennes bio-heat transfer model states that the rate at which heat is stored in soft tissue is due to the conjugated heat transfer of the rate of bio-heat conduction in solid tissue, the rate of convective heat transfer between blood vessels and surrounding tissue, the rate of metabolic heat generation, and the rate of regional heat sources. For a time-continuous three-dimensional (3-D) spatial domain $(x, y, z)$, the strong-form equation governing the Pennes bio-heat transfer is given by

$$\rho c \frac{\partial T(t)}{\partial t} = k\left(\frac{\partial^2 T(t)}{\partial x^2} + \frac{\partial^2 T(t)}{\partial y^2} + \frac{\partial^2 T(t)}{\partial z^2}\right) + w_b c_b \big(T_a - T(t)\big) + Q_m(t) + Q_r(t); \quad t \geq t_0 \tag{1}$$

where $\rho$ is the tissue density $[kg/m^3]$; $c$ the tissue specific heat capacity $[J/(kg \cdot ℃)]$; $T(t)$ the tissue temperature [℃] that varies with time $t$ $[s]$; $k$ the tissue thermal conductivity $[W/(m \cdot ℃)]$; $w_b$ the blood perfusion rate $[kg/(m^3 \cdot s)]$; $c_b$ the blood specific heat capacity $[J/(kg \cdot ℃)]$; $T_a$ the arterial temperature [℃]; $Q_m(t)$ the metabolic heat generation rate $[W/m^3]$; $Q_r(t)$ the regional heat source $[W/m^3]$; $t_0$ the initial time at which the initial condition is $T(t_0) = T_0$ denoting the initial temperature of the tissue.

Equation. (1) presents the Pennes bio-heat equation based on the assumptions of constant soft tissue density $\rho$, tissue specific heat capacity $c$, and tissue thermal conductivity $k$; however, for soft biological tissue, these thermal properties are, in general, vary with temperature [16-18]. By considering the temperature-dependent characteristics of these thermal parameters, Eq. (1) may be written as

$$\rho(T) c(T) \frac{\partial T(t)}{\partial t} = k(T)\left(\frac{\partial^2 T(t)}{\partial x^2} + \frac{\partial^2 T(t)}{\partial y^2} + \frac{\partial^2 T(t)}{\partial z^2}\right) + w_b c_b \big(T_a - T(t)\big) + Q_m(t) + Q_r(t) \tag{2}$$

where $\rho(T)$ is the temperature-dependent tissue density $[kg/m^3]$ in the function of temperature $T$; $c(T)$ the temperature-dependent tissue specific heat capacity $[J/(kg \cdot ℃)]$; and $k(T)$ the temperature-dependent tissue thermal conductivity $[W/(m \cdot ℃)]$.

To fully define the Pennes bio-heat transfer equation, it is necessary to specify the thermal boundary conditions. The thermal boundary $\Gamma$ in the Pennes bio-heat transfer problem can be decomposed into the following three different boundary types: $\Gamma = \Gamma_E + \Gamma_H + \Gamma_A$, and their associated thermal conditions are expressed as follows:

- Essential boundary condition $\Gamma_E$

$$T(\mathbf{x}, t) = T_\Gamma, \quad t > t_0, \mathbf{x} \in \Gamma_E \tag{3}$$

where $T(\mathbf{x}, t)$ is the temperature at spatial point $\mathbf{x}$ and time $t$, and $T_\Gamma$ is the prescribed temperature on the Dirichlet boundary.

- Heat flux boundary condition $\Gamma_H$

$$-k(T)\left(\frac{\partial T(t)}{\partial x} n_x + \frac{\partial T(t)}{\partial y} n_y + \frac{\partial T(t)}{\partial z} n_z\right) = \mathbf{q}(\mathbf{x}, t), \quad t > t_0, \mathbf{x} \in \Gamma_H \tag{4}$$

where $n_x$, $n_y$, and $n_z$ are the components of the outward unit normal direction on the boundary, and $\mathbf{q}(\mathbf{x}, t)$ is the prescribed heat flux on the Neumann boundary.

- Adiabatic boundary condition $\Gamma_A$

$$k(T)\left(\frac{\partial T(t)}{\partial x} n_x + \frac{\partial T(t)}{\partial y} n_y + \frac{\partial T(t)}{\partial z} n_z\right) = 0, \quad t > t_0, \mathbf{x} \in \Gamma_A \tag{5}$$

### *2.2 Fast explicit dynamics finite element algorithm (FED-FEM)*

As mentioned previously, fast and efficient solution methods to bio-heat transfer problems are of great importance in biomedical applications such as the ones for thermo-therapeutic treatment. To this end, the proposed methodology is based on the framework of fast explicit dynamics finite element algorithm (FED-FEM; see Ref. [36] for details) which can achieve fast computation of transient heat transfer and is well suited for real-time thermal applications. The main contributions of FED-FEM for achieving fast run-time computation are:

- explicit dynamics in the temporal domain,
- element-level thermal load computation,
- computationally efficient finite elements,
- explicit formulation for unknown nodal temperature,
- pre-computation of simulation parameters.

The above contributions lead to the following computational benefits of FED-FEM:

- eliminating the need for inversion of thermal stiffness matrix for time stepping,
- eliminating the need for assembling thermal stiffness matrix of the entire model,
- obtaining the thermal responses efficiently in the spatial domain,
- eliminating the need for iterations anywhere in the algorithm and permitting parallel implementation,
- facilitating computational efficiency of run-time computation.

The FED-FEM allows straightforward treatment of nonlinearities and can accommodate non/linear thermal properties and boundary conditions. Different from standard FEM, the computations are performed at the element level, and a time step is performed directly without the need for forming the global system of equations and applying numerical iterations for solution. The computation cost at each time step is low, and the independent equations for individual nodal unknown temperature can be computed independently, permitting parallel implementation of the proposed methodology for fast solutions of bio-heat transfer problems. These key features make the proposed methodology well suited for real-time computations.

### *2.3 FED-FEM formulation of Pennes bio-heat transfer model*

#### *2.3.1 Explicit dynamics in the temporal domain*

To achieve fast solutions of the Pennes bio-heat transfer equation in the temporal domain, it requires an efficient time-stepping scheme for temporal integrations at each time step during the simulation. The explicit integration [44, 45] is computationally efficient compared to the computationally more intensive implicit integration [9, 46]. It uses the forward finite difference scheme where field variables in the future state are obtained explicitly based on the current state of known values only, without the need for inversion of the system stiffness matrix at each time step [45]; hence, it is easy to implement and computationally efficient. The explicit integration is also well suited for parallel computing since mass lumping (diagonal mass) can be employed through which the global system of equations can be split into independent equations for individual nodes, allowing each nodal equation to be assigned to a computation in the parallel processor to perform calculations independently.

Using the first-order explicit forward time integration scheme, the temporal derivative of the continuous field temperature is written as

$$\frac{\partial T(t)}{\partial t} = \frac{T(t+\Delta t) - T(t)}{\Delta t} \tag{6}$$

where $\Delta t$ is the time step.

By substituting Eq. (6) into the spatially discretised matrix form of the Pennes bio-heat transfer equation, yielding

$$\mathbf{C}(T)\left(\frac{\mathbf{T}(t+\Delta t) - \mathbf{T}(t)}{\Delta t}\right) = \mathbf{K}(T)\mathbf{T}(t) - \mathbf{K}_b\mathbf{T}(t) + \mathbf{Q}_b + \mathbf{Q}_m(t) + \mathbf{Q}_r(t) \tag{7}$$

where $\mathbf{C}(T)$ is the thermal mass (tissue density and specific heat) matrix; $\mathbf{T}(t)$ the vector of nodal temperatures at time $t$; $\mathbf{K}(T)$ the thermal stiffness (conduction) matrix; $\mathbf{K}_b$ the thermal stiffness (blood perfusion) matrix; $\mathbf{Q}_b$ the vector of heat flows of blood perfusion at arterial temperature; $\mathbf{Q}_m(t)$ the vector of heat flows of metabolic heat generation; and $\mathbf{Q}_r(t)$ the vector of heat flows of regional heat sources.

Equation. (7) may be further arranged into

$$\mathbf{T}(t+\Delta t)=\frac{\Delta t}{\mathbf{C}(T)}\big(\mathbf{K}(T)\mathbf{T}(t)-\mathbf{K}_b\mathbf{T}(t)+\mathbf{Q}_b+\mathbf{Q}_m(t)+\mathbf{Q}_r(t)\big)+\mathbf{T}(t) \tag{8}$$

Since the explicit scheme is conditionally stable, precautions need to be taken for the time increment steps to meet the Courant-Friedrichs-Lewy (CFL) condition [47] for numerical stability. The numerical stability is assured only for time step size $\Delta t<\Delta t_{cr}$ where the critical time increment $\Delta t_{cr}$ is determined by [44]

$$\Delta t_{cr}=\frac{2}{|\lambda_{max}|} \tag{9}$$

where $\lambda_{max}$ is the largest eigenvalue of the global thermal stiffness over global thermal mass values [44, 45, 48]; however, it is worth noting that the computation of eigenvalues of heat transfer problems is computationally very expensive, especially when the thermal matrices are large. Some methods for prediction of the largest eigenvalues and critical time steps were reported [44], and the numerical stability in dynamic heat transfer was also analysed [49].

### *2.3.2 Element-level thermal load computation*

Equation. (8) may be further written in terms of element-level thermal conduction loads, yielding

$$\mathbf{T}(t+\Delta t)=\frac{\Delta t}{\mathbf{C}(T)}\left(\sum_e \mathbf{F}_e(T,t)-\mathbf{K}_b\mathbf{T}(t)+\mathbf{Q}_b+\mathbf{Q}_m(t)+\mathbf{Q}_r(t)\right)+\mathbf{T}(t) \tag{10}$$

where

$$\sum_e \mathbf{F}_e(T,t)=\mathbf{K}(T)\mathbf{T}(t)=\mathbf{F}(T,t) \tag{11}$$

where $\mathbf{F}_e(T,t)$ are the components due to bio-heat conduction in element $e$ of the global nodal thermal loads $\mathbf{F}(T,t)$. For a given element $e$, $\mathbf{F}_e(T,t)$ may be computed by

$$\mathbf{F}_e(T,t)=\int_{V_e} k(T)\mathbf{B}(\mathbf{x})^T\mathbf{B}(\mathbf{x})\,dV\,\mathbf{T}(t) \tag{12}$$

where $V_e$ is the $e$th element volume, and $\mathbf{B}(\mathbf{x})$ the temperature gradient matrix.

Due to the temperature-dependent property of the element thermal conductivity $k(T)$, its value is determined by calculating the temperature-dependent thermal conductivity at element nodes using nodal temperatures and averaging over nodal values, yielding

$$k(T)=\frac{1}{N}\sum_{n=1}^{N}k_n(T) \tag{13}$$

where $k_n(T)$ is the thermal conductivity at the $n$th node in the element.

Equations. (10-12) imply that computations are performed at the element level, eliminating the need for assembling the global thermal stiffness matrix $\mathbf{K}(T)$ and multiplying the temperature $\mathbf{T}(t)$ for the entire model. Therefore, the computational cost at each time step is significantly lower in the proposed methodology compared to standard FEM since there is no need for forming the global system of equations and applying iterations for the solution.

### *2.3.3 Computationally efficient finite elements*

The computation of the integral in Eq. (12) must be conducted at every integration point for element nodal thermal loads. To achieve fast and efficient computation, the computationally efficient 3-D low-order finite elements such as the eight-node reduced integration hexahedral element and the four-node linear tetrahedral element are used for 3-D Pennes bio-heat transfer problems.

For the eight-node reduced integration hexahedral element, it leads to the formulation for element nodal thermal loads, yielding

$$\mathbf{F}_e(T,t) = k(T)8\det(\mathbf{J})\mathbf{B}(\mathbf{x})^T\mathbf{B}(\mathbf{x})\mathbf{T}(t) \tag{14}$$

where 8 is the constant integer for volume integral of the eight-node reduced integration hexahedral element (see Ref. [48] for details), and $\mathbf{J}$ the pre-computed element Jacobian matrix defining the mapping between derivatives in global and element natural coordinates.

For the four-node linear tetrahedral element, it leads to the formulation for element nodal thermal loads, yielding

$$\mathbf{F}_e(T,t) = k(T)V_{tet}\mathbf{B}(\mathbf{x})^T\mathbf{B}(\mathbf{x})\mathbf{T}(t) \tag{15}$$

where $V_{tet}$ is the volume of the tetrahedral element.

By defining the matrix $\mathbf{A}(\mathbf{x})$

$$\mathbf{A}(\mathbf{x}) = \begin{cases} 8\det(\mathbf{J})\mathbf{B}(\mathbf{x})^T\mathbf{B}(\mathbf{x}) & \text{reduced integration hexahedrons} \\ V_{tet}\mathbf{B}(\mathbf{x})^T\mathbf{B}(\mathbf{x}) & \text{linear tetrahedrons} \end{cases} \tag{16}$$

the element nodal thermal loads can be computed as

$$\mathbf{F}_e(T,t) = k(T)\mathbf{A}(\mathbf{x})\mathbf{T}(t) \tag{17}$$

Since the temperature gradient matrix $\mathbf{B}(\mathbf{x})$, the element Jacobian matrix $\mathbf{J}$, and the volume of a tetrahedral element $V_{tet}$ may be pre-computed, the matrix $\mathbf{A}(\mathbf{x})$ may also be pre-computed before the commencement of run-time simulation; hence, at each time step the nodal thermal loads are updated by considering only the temperature-dependent thermal conductivity $k(T)$ and variations of nodal temperatures $\mathbf{T}(t)$.

#### *2.3.4 Explicit formulation for unknown nodal temperature*

By employing the lumped (diagonal) mass approximation in which the mass is distributed equally among the element nodes while ensuring the conservation of mass, it leads to a diagonal thermal mass matrix and thermal stiffness (blood perfusion) matrix, rendering Eq. (10) an explicit formulation for the unknown field temperatures $\mathbf{T}(t+\Delta t)$ at the next time point. Assuming that the nodal thermal loads $\mathbf{F}(T,t)$ have been computed using Eq. (11), the explicit forward time integration renders an equation for obtaining nodal temperatures at the next time point, yielding

$$\mathbf{T}(t+\Delta t) = \frac{\Delta t}{\mathbf{C}(T)}\left(\mathbf{F}(T,t) - \mathbf{K}_b\mathbf{T}(t) + \mathbf{Q}_b + \mathbf{Q}_m(t) + \mathbf{Q}_r(t)\right) + \mathbf{T}(t) \tag{18}$$

where $\mathbf{C}(T)$ and $\mathbf{K}_b$ are the lumped thermal mass and stiffness (blood perfusion) matrices.

For each time step, the field temperature at node $i$ can be obtained by

$$T^{(i)}(t+\Delta t) = \left(\frac{\Delta t}{C^{(i)}(T)}\right)\left(F^{(i)}(T,t) - {K_b}^{(i)}T^{(i)}(t) + {Q_b}^{(i)} + {Q_m}^{(i)}(t) + {Q_r}^{(i)}(t)\right) + T^{(i)}(t) \tag{19}$$

where $C^{(i)}(T)$ and ${K_b}^{(i)}$ are the diagonal entries in the $i$th row of the diagonal thermal mass matrix $\mathbf{C}(T)$ and diagonal thermal stiffness (blood perfusion) matrix $\mathbf{K}_b$, respectively.

The temperature-dependent thermal properties are accounted for in the calculations of thermal mass matrix $\mathbf{C}(T)$ and the nodal thermal loads $\mathbf{F}(T,t)$, and nonlinear thermal boundary conditions are accounted for in the calculations of the nodal heat flows $\mathbf{Q}$. It is worth noting that there is no need for iterations anywhere in the algorithm; hence, Eq. (19) provides an efficient means for advancing temperatures in time. Furthermore, it also states an explicit formulation of the unknown field temperatures $\mathbf{T}(t+\Delta t)$; therefore, the global system of equations can be split into independent equations for individual nodes, permitting parallel implementation of the proposed algorithm to perform calculations independently.

*2.3.5 Pre-computation of simulation parameters*

Since most studies on the Pennes bio-heat transfer equation employed temperature-independent blood perfusion rate $w_b$, blood specific heat capacity $c_b$, and arterial temperature $T_a$ [20, 21, 30, 40, 41, 50, 51], and it is considered in this work that only the tissue density $\rho(T)$, specific heat capacity $c(T)$, and thermal conductivity $k(T)$ are temperature-dependent variables; therefore, similar to the matrix $\mathbf{A}(\mathbf{x})$ which can be pre-computed, the diagonal thermal stiffness (blood perfusion) matrix $\mathbf{K}_b$ and the vector of heat flows of blood perfusion at arterial temperature $\mathbf{Q}_b$ may also be pre-computed to facilitate the computational performance at run-time computation.

*2.4 Parallel implementation*

It can be seen that Eq. (19) presents an explicit formulation to advance nodal temperatures in the temporal domain where nodal temperatures at the next time point are solely computed from the variables at the current time point of known values. Furthermore, the system equation of the soft tissue model is represented by individual nodal equations, leading to a system of uncoupled equations where no assembly of the global system of equations is required. This is particularly suitable for distributed parallel computing where each nodal equation can be assigned to a computation in the parallel processor to perform calculations independently. In addition, the nodal thermal load contributions are computed from the finite elements, and hence each element-level computation can be also assigned to a computation in the parallel processor to perform calculations independently. Also, the parallel implementation of FED-FEM has not yet demonstrated in [36]; therefore, the parallel implementation is demonstrated in the present work for the Pennes bio-heat transfer problem. The algorithm of the proposed methodology is presented in Algorithm. 1.

Algorithm 1 Simulation algorithm

Pre-computation

1 : load soft tissue mesh and boundary conditions

2 : **for all** elements **do**

3 : compute temperature gradient matrix $\mathbf{B}(\mathbf{x})$, determinant of element Jacobian $\det(\mathbf{J})$, and tetrahedral element volume $V_{tet}$

4 : compute constant matrix $\mathbf{A}(\mathbf{x})$ using Eq. (16)

5 : compute diagonal (constant) thermal stiffness (blood perfusion) matrix $\mathbf{K}_b$ and the vector of heat flows of blood perfusion at arterial temperature $\mathbf{Q}_b$

6 : **end for**

Initialisation

7 : Initialise nodal temperature $T_0$ at $t = t_0$, and apply thermal boundary condition $\Gamma_{H,\Delta t}$ for the first time step $\Delta t$

Time-stepping

8 : **parallel for all** elements **do**

9 : compute element thermal loads $\mathbf{F}_e(T, t)$ using Eq. (17)

10 : add to global nodal thermal loads $\mathbf{F}(T, t) \leftarrow \mathbf{F}_e(T, t)$ using Eq. (11)

11 : **end for**

12 : **parallel for all** nodes **do**

13 : compute $T(t + \Delta t)$ using Eq. (19)

14 : **end for**

15 : update $\Gamma_{H,t+\Delta t}$ for next time step $t + \Delta t$

## 3. Results

### *3.1 Algorithm verification*

The proposed methodology is compared against the commercial finite element analysis package, ABAQUS/CAE 2018 (License 6.20) to verify the numerical accuracy before evaluating the computational performance. It should be mentioned that the numerical convergence of FED-FEM has been verified via the standard patch tests [36]. Numerical accuracy of the individual terms of the Pennes bio-heat transfer equation such as the tissue bio-heat conduction, blood perfusion, and metabolic heat generation are evaluated individually before evaluating the combined effect as a whole, and assessments of numerical errors are presented. Furthermore, the numerical accuracy of the proposed method for temperature-dependent nonlinear thermal properties is also investigated against the ABAQUS solution.

#### *3.1.1 Model setup*

As illustrated in Fig. 1, the object under investigation is a virtual model of the human liver, and it is discretised into a tetrahedral mesh of four-node linear heat transfer tetrahedrons, leading to 13170 nodes and 64870 elements. Soft tissue thermal material properties are the constant isotropic and homogeneous thermal conductivity $k = 0.518\ W/(m \cdot {}^{\circ}\mathrm{C})$, tissue density $\rho = 1060\ kg/m^3$, and specific heat $c = 3700\ J/(kg \cdot {}^{\circ}\mathrm{C})$ [52]. The initial tissue temperature is $T_0 = 37\ {}^{\circ}\mathrm{C}$. The nodes at the bottom of the liver are prescribed with a constant temperature $T_\Gamma = 37\ {}^{\circ}\mathrm{C}$ throughout the simulation.

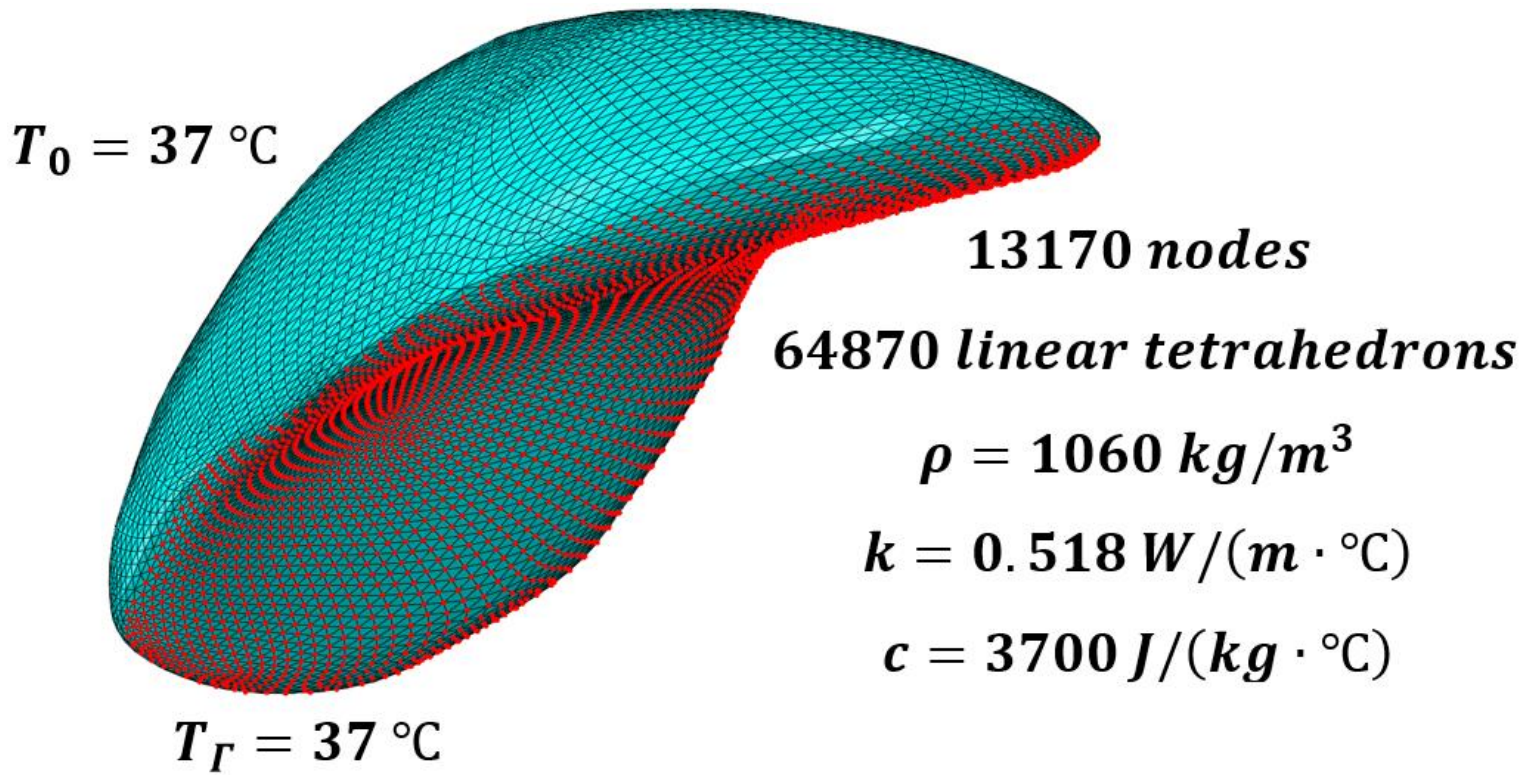

Fig. 1. Geometry, initial, and boundary conditions of the tested liver model: the virtual human liver model is discretised into a linear tetrahedral mesh consisting of 13170 nodes and 64870 elements; the initial tissue temperature is $T_0 = 37\ {}^{\circ}\mathrm{C}$ at all nodes; the nodes (shown in red) at the bottom side of the liver are prescribed with a constant temperature $T_\Gamma = 37\ {}^{\circ}\mathrm{C}$ throughout the simulation.

#### *3.1.2 Verification of bio-heat conduction*

As illustrated in Fig. 2, the bio-heat conduction in the Pennes bio-heat transfer equation is simulated by the proposed methodology, and its numerical accuracy is compared against ABAQUS solutions by applying a constant concentrated heat flux $q = 2\ W$ to the red nodes (see Fig. 2(a)). The adiabatic boundary condition is applied to the remaining nodes. For numerical comparison, the temperature-history at the five red nodes (see Fig. 2(b)) are monitored throughout the simulation. The simulation is conducted for $2000$ time steps for a duration of $t = 10\ s$ bio-heat conduction with a time step size $\Delta t = 0.005\ s$, and it can achieve stable simulation. It can be seen from Fig. 2(e-f) that there is a good agreement with those from the proposed methodology (P,N#) compared to the ABAQUS solution (A,N#) with the same set of mesh and simulation parameters.

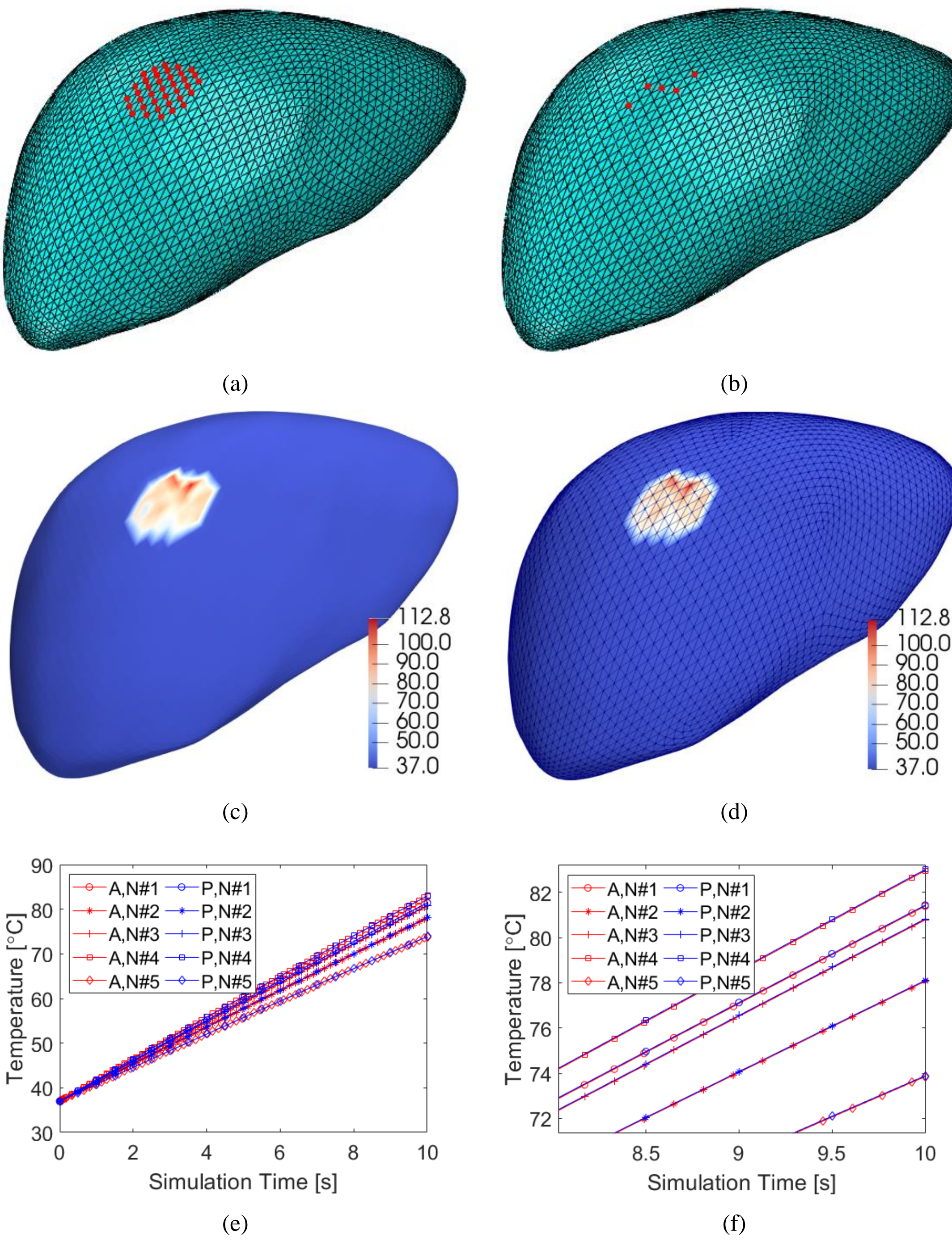

Fig. 2. Loading condition and simulation results: (a) the red nodes are applied with a constant concentrated heat flux $q = 2\,W$; (b) temperature-history at the five red nodes are monitored for numerical comparison; (c) temperature distribution at the completion $t = 10\ s$; (d) same temperature distribution with tetrahedral mesh displayed; (e) comparison of temperature-history between the results of the proposed methodology (P,N#) and ABAQUS (A,N#) for the five red nodes (#1-5); and (f) a close-up look of the temperature-history.

At the completion of simulation $t = 10\ s$, the statistical results of the comparison at all nodes are presented in Table. 1. It can be seen that there is only a marginal difference between the results of the proposed methodology and ABAQUS solutions. The numerical error used in the present work is calculated using the following equation:

$$Error = \sqrt{\frac{\sum_{n=1}^{N}\left(T_n^{ABAQUS} - T_n^{Proposed}\right)^2}{\sum_{n=1}^{N}\left(T_n^{Abaqus}\right)^2}} \tag{20}$$

Table. 1

Statistical results for all nodes at the completion $t = 10\ s$: A (ABAQUS), P (Proposed), D (Difference), and E (Error).

| | Min (°C) | Max (°C) | Median (°C) | RMS (°C) | Q1 (°C) | Q3 (°C) |
|---|---|---|---|---|---|---|
| A | 36.7093 | 112.6180 | 37 | 37.2120 | 37 | 37 |
| P | 36.7119 | 112.7800 | 37 | 37.2123 | 37 | 37 |
| D | -0.1620 | 0.0314 | 0 | 0.0037 | 0 | 0 |
| E | | | 1.0072e-04 | | | |

### *3.1.3 Verification of blood perfusion*

The verification of simulating blood perfusion in the Pennes bio-heat transfer equation is conducted by applying a constant concentrated film condition $h = 0.0293\ W/(m^2 \cdot °C)$ to all nodes in the liver model with an associated nodal area of $1\ m^2$. The employed $h$ value is determined based on the blood perfusion rate $w_b = 26.6\ kg/(m^3 \cdot s)$, blood specific heat $c_b = 3617\ J/(kg \cdot °C)$ [52], and liver volume $V = 0.00401\ m^3$ of the model. The arterial temperature is set to $T_a = 39$ °C. The simulation is conducted for $3000$ time steps for a duration of $t = 30\ s$ blood perfusion with a time step size $\Delta t = 0.01\ s$, and it can achieve stable simulation. Fig. 3 illustrates the temperature-history at the monitored nodes and simulation results. It can be seen from Fig. 3(c-d) there is a good agreement with those from the proposed methodology (P,N#) compared to the ABAQUS solution (A,N#). At the completion $t = 30\ s$, the statistical results of the comparison at all nodes are presented in Table. 2.

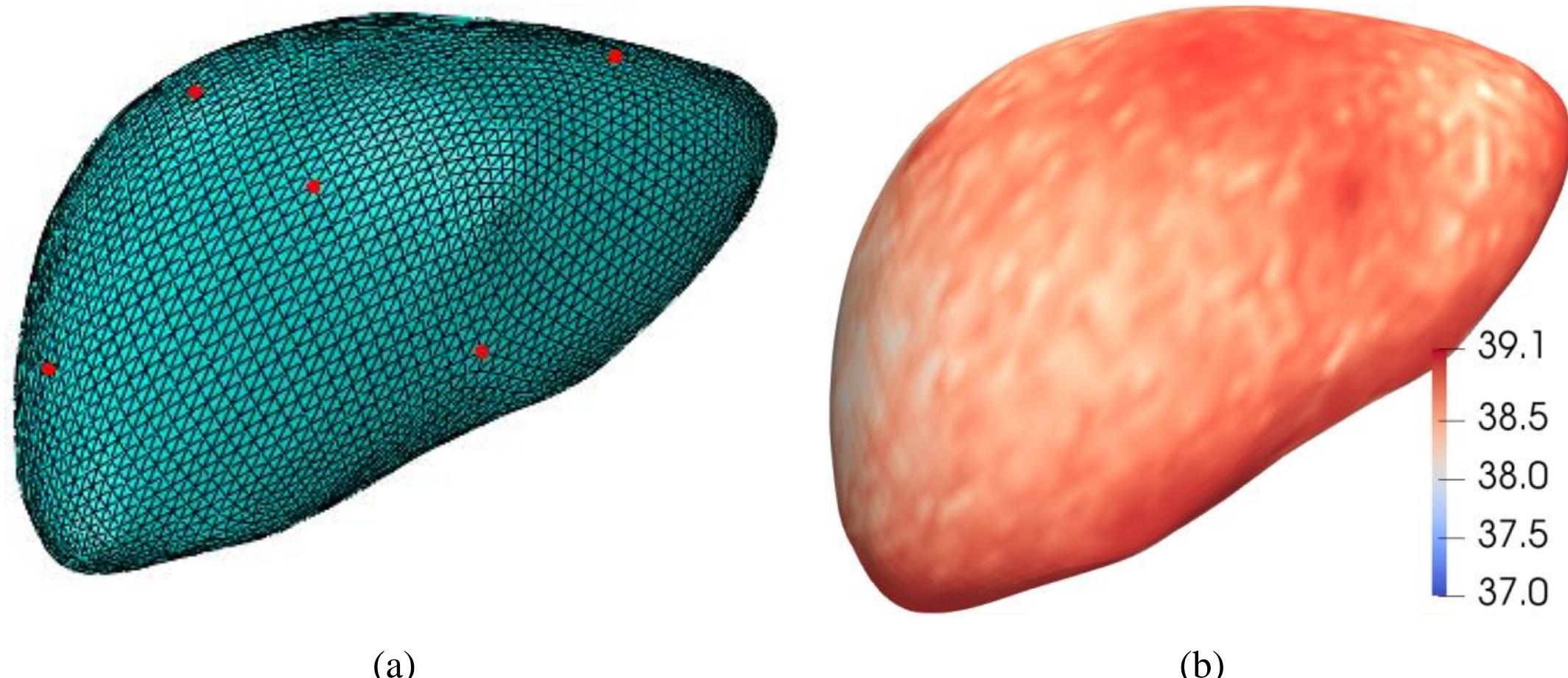

(a) (b)

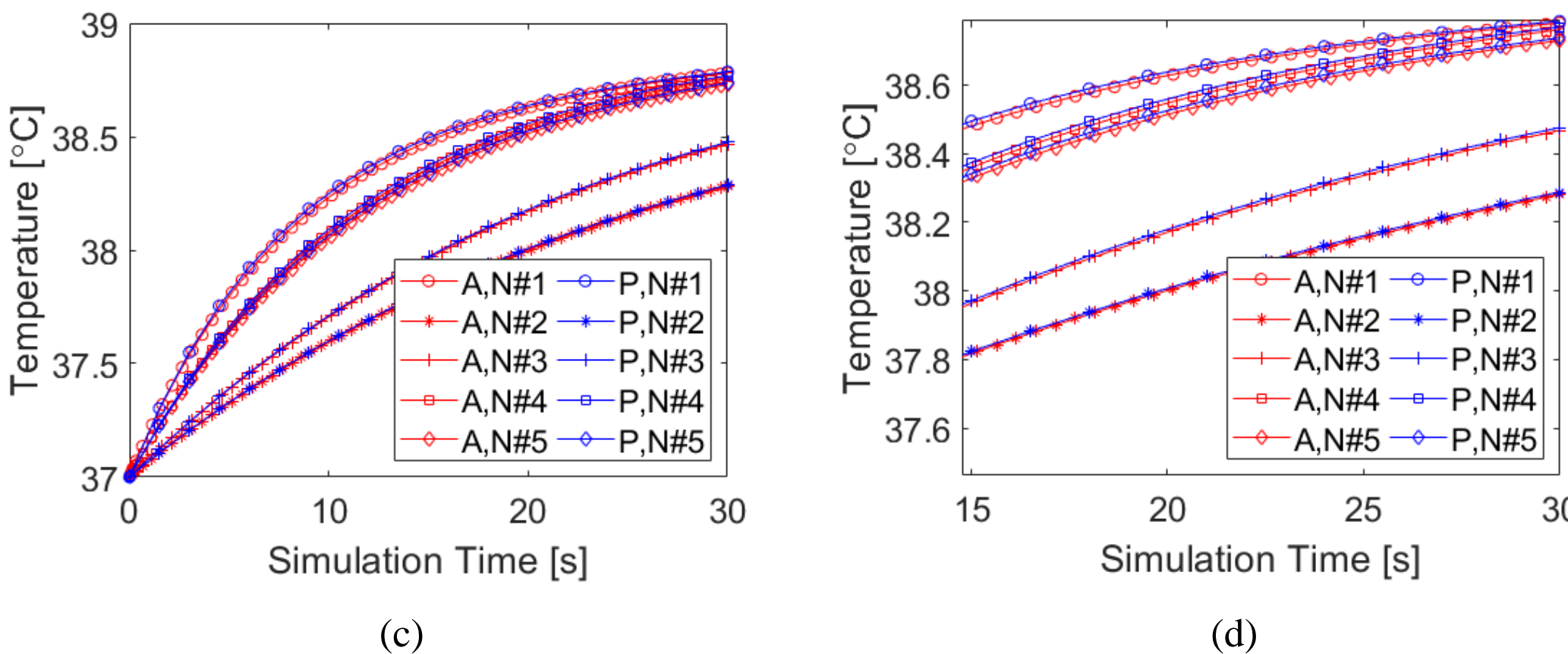

(c) (d)

Fig. 3. Temperature-history at the monitored nodes and simulation results: (a) temperatures at the five red nodes are monitored for numerical comparison; (b) temperature distribution at completion $t = 30\ s$; (c) comparison of temperature-history between the results of the proposed method (P,N#) and ABAQUS (A,N#) for the five red nodes (#1-5); and (d) a close-up look of the temperature-history.

Table. 2

Statistical results for all nodes at the completion $t = 30\ s$: A (ABAQUS), P (Proposed), D (Difference), and E (Error).

| | Min (℃) | Max (℃) | Median (℃) | RMS (℃) | Q1 (℃) | Q3 (℃) |
|---|---|---|---|---|---|---|
| A | 37 | 39.1077 | 38.1160 | 38.0118 | 37.3791 | 38.6156 |
| P | 37 | 39.1076 | 38.1214 | 38.0157 | 37.3796 | 38.6243 |
| D | -0.0102 | 0.0046 | -0.0037 | 0.0050 | -0.0069 | -0.0005 |
| E | 1.3152e-04 | | | | | |

*3.1.4 Verification of metabolic heat generation*

The proposed methodology for simulating metabolic heat generation in the Pennes bio-heat transfer equation is verified by applying a constant concentrated heat flux $q = 0.0103\ W$ to all nodes in the liver model. The applied concentrated heat flux is assumed to be a common value at all nodes for the purpose of verification of numerical accuracy compared to ABAQUS solutions. The employed value for the concentrated heat flux is determined based on the metabolic heat generation rate $Q_m = 33800\ W/m^3$ [51] and liver volume $V = 0.00401\ m^3$ for the 13170 nodes. The simulation is conducted for 2000 time steps for a duration of $t = 20\ s$ metabolic heat generation with a time step size $\Delta t = 0.01\ s$, and it can achieve stable simulation. Fig. 4 illustrates the temperature-history at the monitored nodes and simulation results. It can be seen from Fig. 4(c-d) there is a good agreement with those from the proposed methodology (P,N#) compared to the ABAQUS solution (A,N#). At the completion $t = 20\ s$, the statistical results of the comparison at all nodes are presented in Table. 3.

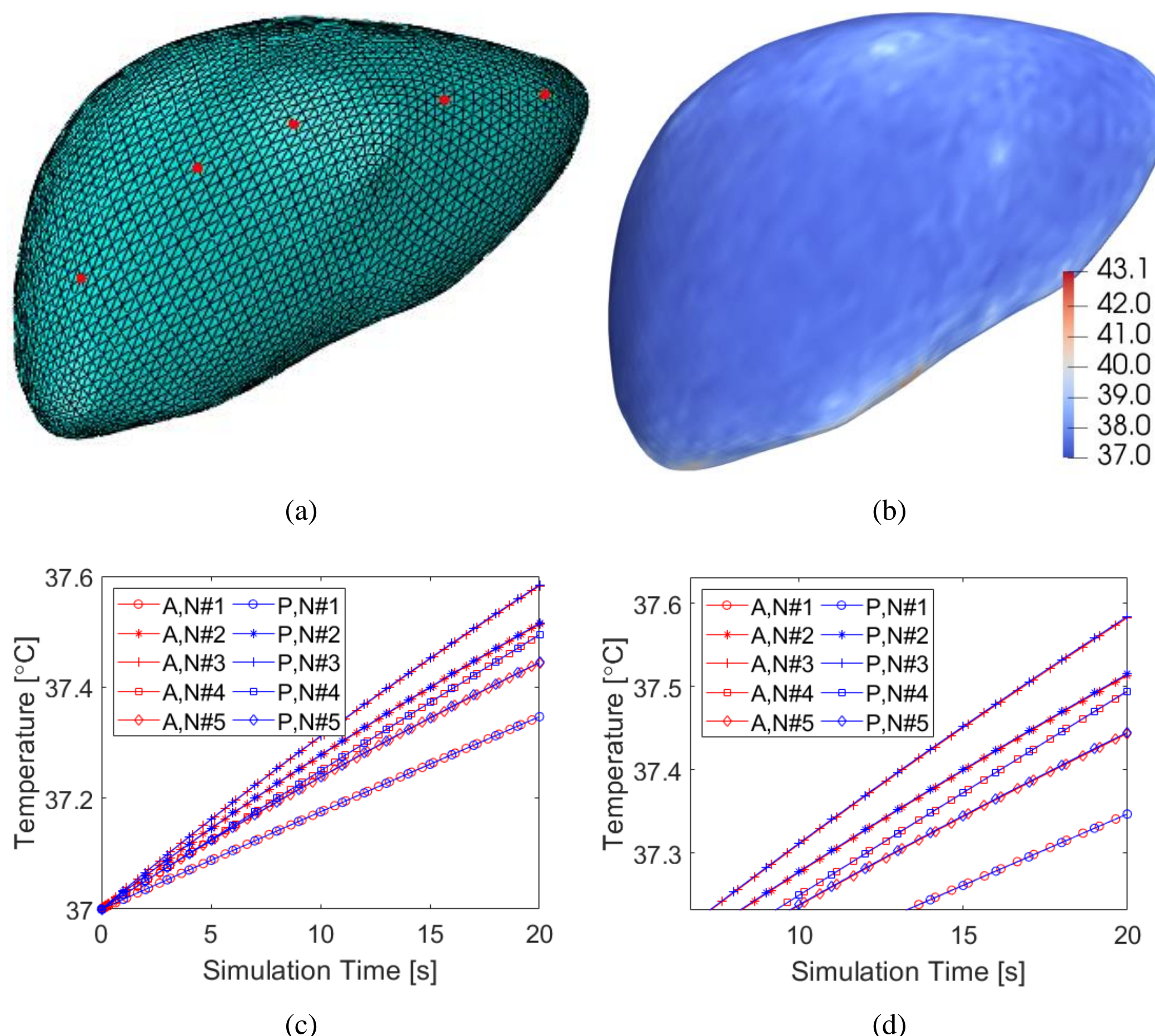

Fig. 4. Temperature-history at the monitored nodes and simulation results: (a) temperatures at the five red nodes are monitored for numerical comparison; (b) temperature distribution at completion $t = 20\ s$; (c) comparison of temperature-history between the results of the proposed method (P,N#) and ABAQUS (A,N#) for the five red nodes (#1-5); and (d) a close-up look of the temperature-history.

Table. 3

Statistical results for all nodes at the completion $t = 20\ s$: A (ABAQUS), P (Proposed), D (Difference), and E (Error).

| | Min (℃) | Max (℃) | Median (℃) | RMS (℃) | Q1 (℃) | Q3 (℃) |
|---|---|---|---|---|---|---|
| A | 37 | 43.1066 | 37.1975 | 37.3442 | 37.0490 | 37.4327 |
| P | 37 | 43.1352 | 37.1980 | 37.3449 | 37.0485 | 37.4334 |
| D | -0.0332 | 0.0034 | -0.0001 | 0.0021 | -0.0009 | 0 |
| E | 5.6936e-05 | | | | | |

*3.1.5 Verification of combined effect (Pennes bio-heat transfer)*

The combined effect of bio-heat conduction, blood perfusion, and metabolic heat generation is verified in this section using the proposed methodology. The verifications of the bio-heat conduction, blood perfusion, and metabolic heat generation have been demonstrated previously in Sections 3.1.2-4, respectively.

The Pennes bio-heat transfer equation is verified by applying a two-stage simulation process: (i) a $t = 3\ s$ thermal heating process where the liver is exposed to regional heat sources illustrated in Section 3.1.2, and (ii) a $t = 17\ s$ relaxation process where the regional heat sources are removed. Soft tissue thermal material properties for bio-heat conduction, blood perfusion, and metabolic heat generation in Sections 3.1.2-4 are again employed. The arterial temperature is set to $T_a = 37$ ℃. The simulation is conducted for $2000$ time steps for a duration of $t = 3\ s$ thermal heating and $t = 17\ s$ relaxation with a time step size $\Delta t = 0.01\ s$, and it can achieve stable simulation. Fig. 5 presents the temperature-history at the monitored nodes and simulation results. It can be seen from Fig. 5(c-d) there is a good agreement with those from the proposed methodology (P,N#) compared to the ABAQUS solution (A,N#). At the completion $t = 20\ s$, the statistical results of the comparison at all nodes are presented in Table. 4.

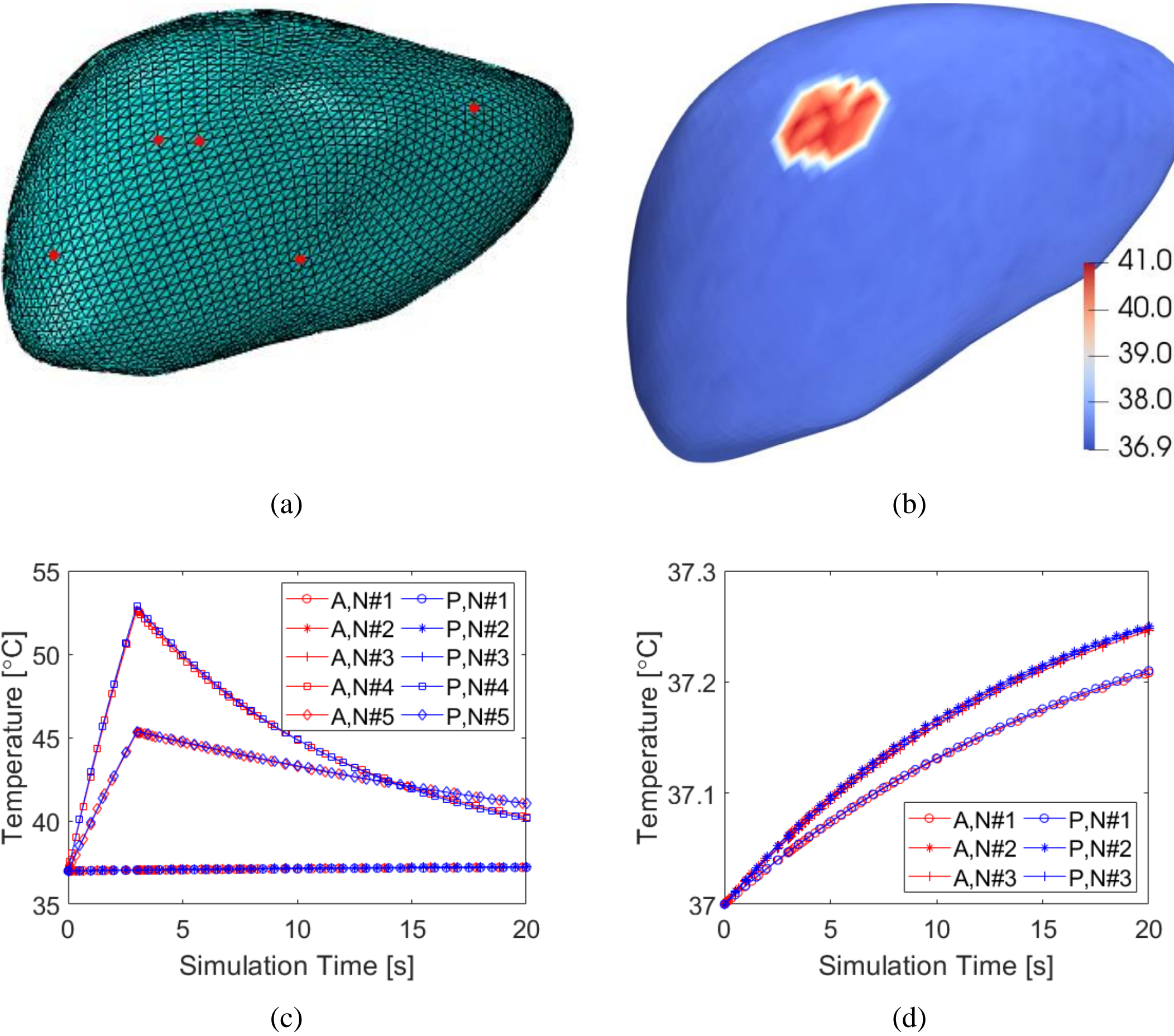

Fig. 5. Temperature-history at the monitored nodes and simulation results: (a) temperatures at the five red nodes are monitored for numerical comparison; (b) temperature distribution at completion $t = 20\ s$; (c) comparison of temperature-history between the results of the proposed methodology (P,N#) and ABAQUS (A,N#) for the five red nodes (#1-5); and (d) a close-up look of the temperature-history at nodes #1-3.

Table. 4

Statistical results for all nodes at the completion $t = 20\ s$: A (ABAQUS), P (Proposed), D (Difference), and E (Error).

| | Min (℃) | Max (℃) | Median (℃) | RMS (℃) | Q1 (℃) | Q3 (℃) |
|---|---|---|---|---|---|---|
| A | 36.9519 | 41.0405 | 37.1491 | 37.1597 | 37.0454 | 37.2433 |
| P | 36.9498 | 41.0269 | 37.1502 | 37.1606 | 37.0458 | 37.2456 |
| D | -0.0144 | 0.1542 | -0.0009 | 0.0046 | -0.0021 | 0 |
| E | 1.2355e-04 | | | | | |

### *3.1.6 Temperature-dependent tissue density, thermal conductivity, and specific heat capacity*

The proposed methodology is also verified against ABAQUS solutions by considering nonlinear thermal properties, such as the temperature-dependent tissue density, thermal conductivity, and specific heat capacity for bio-heat transfer problems [17]. Soft tissue thermal material properties are the temperature-dependent homogeneous tissue density $\rho(T) = 1040\ kg/m^3$ @ 37 ℃ and $\rho(T) = 1000\ kg/m^3$ @ 65 ℃, isotropic thermal conductivity $k(T) = 0.53\ W/(m \cdot ℃)$ @ 37 ℃ and $k(T) = 0.57\ W/(m \cdot ℃)$ @ 65 ℃, specific heat capacity $c(T) = 3600\ J/(kg \cdot ℃)$ @ 37 ℃ and $c(T) = 3800\ J/(kg \cdot ℃)$ @ 65 ℃ [17]. A linear interpolation between points is employed to determine the temperature-dependent tissue density $\rho(T)$, thermal conductivity $k(T)$, and specific heat $c(T)$, i.e., $\Delta\rho(T) = -1.4286\ kg/m^3$ per 1 ℃, $\Delta k(T) = 0.0014286\ W/(m \cdot ℃)$ per 1 ℃, and $\Delta c(T) = 7.14286\ J/(kg \cdot ℃)$ per 1 ℃. Using the same two-stage simulation procedure in Section 3.1.5, numerical analyses are conducted for the temperature-independent (TI) thermal properties at 37 ℃ and temperature-dependent (TD) thermal properties. Fig. 6 presents the temperature-history at the five monitored nodes for the TI and TD cases. There is a good agreement with those from the proposed methodology (P,N#) compared to the ABAQUS solution (A,N#). Fig. 7 further investigates the effect of the TD case on the variations of tissue temperature compared to the TI case, where temperature differences are measured by $T_{TD} - T_{TI}$. It can be seen that the proposed methodology can accurately capture the temperature-dependent effect of the TD case. Furthermore, the statistical results of the comparison at all nodes are presented in Table. 5 at the completion $t = 20\ s$.

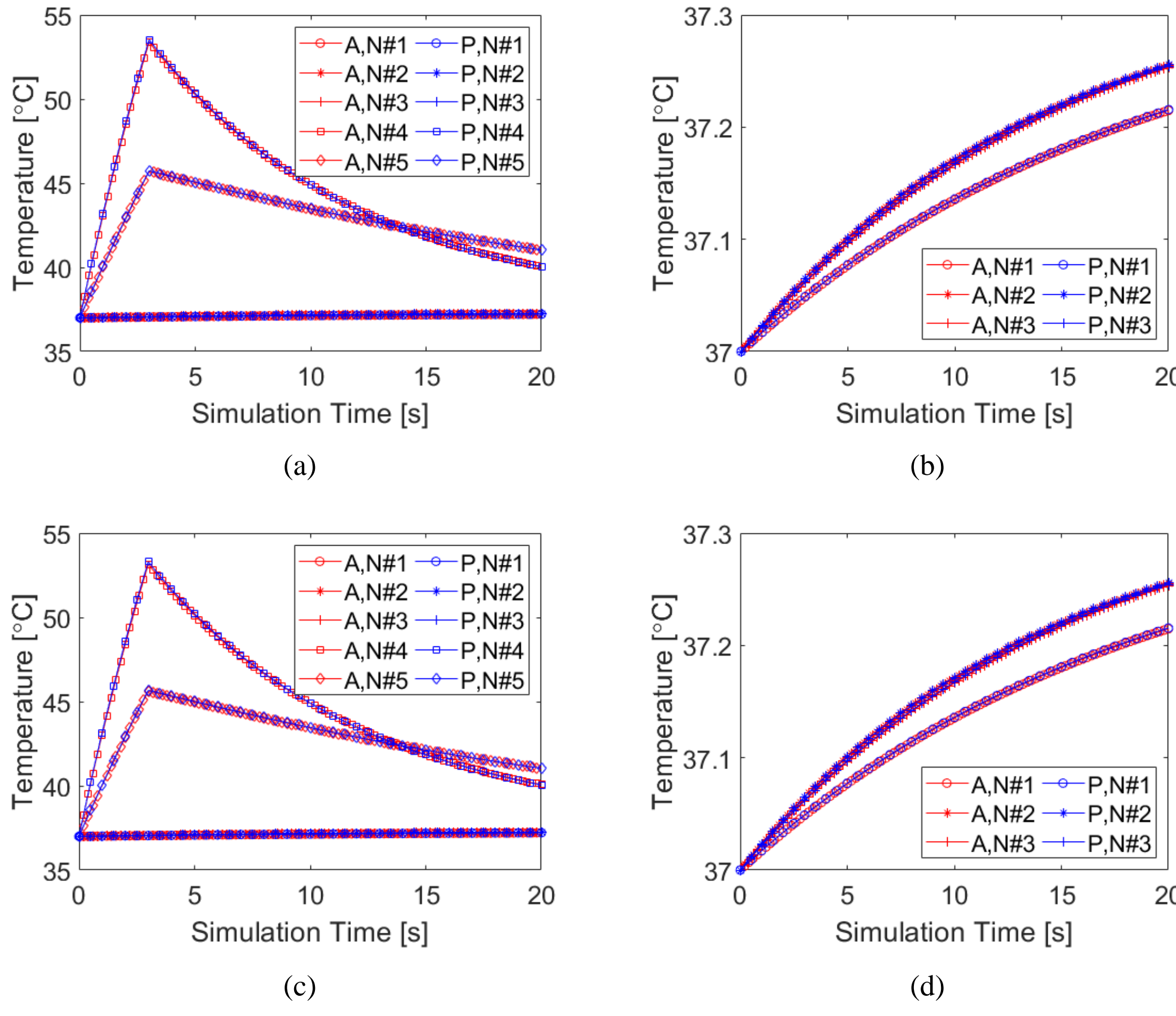

Fig. 6. Comparison of temperature-history between the results of the proposed methodology (P,N#) and ABAQUS (A,N#) for the five red nodes (#1-5) for: (a) TI case; (b) a close-up look of the TI temperature-history at nodes #1-3; (c) TD case; (b) a close-up look of the TD temperature-history at nodes #1-3.

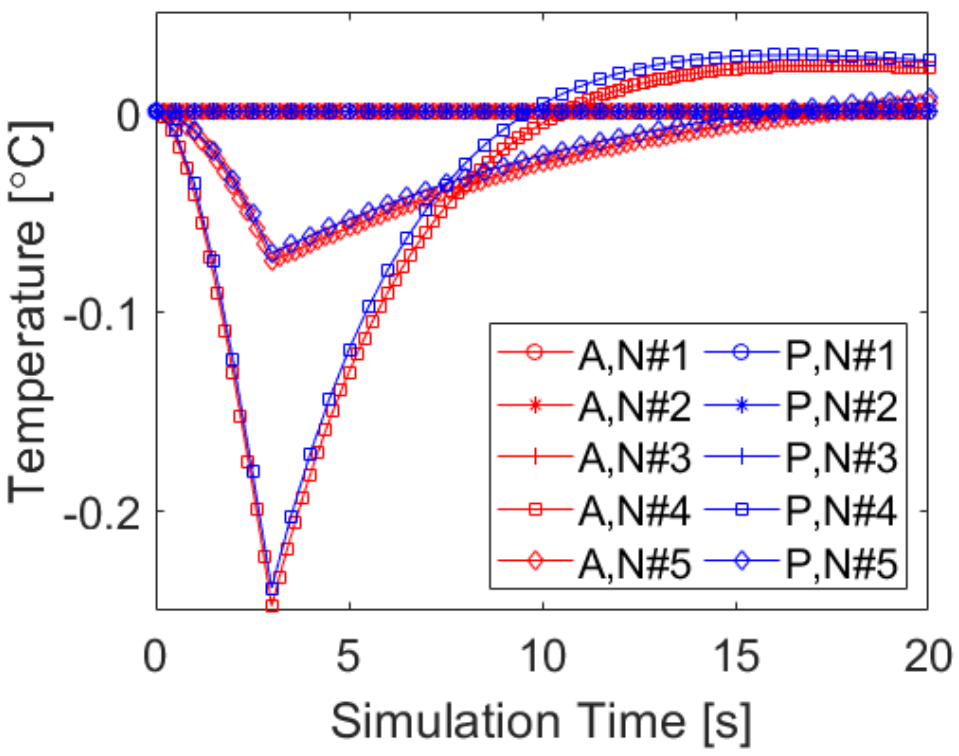

Fig. 7. Comparison of temperature-difference history ($T_{TD} - T_{TI}$) between the results of the proposed methodology (P,N#) and ABAQUS (A,N#) for the five red nodes (#1-5).

Table. 5

Statistical results of the TD case for all nodes at the completion $t = 20\ s$: A (ABAQUS), P (Proposed), D (Difference), and E (Error).

| | Min (°C) | Max (°C) | Median (°C) | RMS (°C) | Q1 (°C) | Q3 (°C) |
|---|---|---|---|---|---|---|
| A | 36.9480 | 41.0330 | 37.1550 | 37.1632 | 37.0476 | 37.2498 |
| P | 36.9470 | 41.0414 | 37.1553 | 37.1636 | 37.0465 | 37.2505 |
| D | -0.0110 | 0.0165 | -0.0003 | 7.5522e-04 | -0.0007 | 0 |
| E | 2.0322e-05 | | | | | |

*3.2 Computational performance*

The computational benefits of the proposed methodology are examined. The proposed method is implemented in C++ and evaluated on an Intel(R) Core(TM) i7-6700 CPU @ 3.40 GHz and 32 GB RAM PC. Six mesh densities, (i, 1445 nodes, 6738 elements), (ii, 1840 nodes, 8724 elements), (iii, 2644 nodes, 12852 elements), (iv, 3339 nodes, 16371 elements), (v, 5304 nodes, 26550 elements), and (vi, 7872 nodes, 40021 elements), are used to evaluate the computation time of the proposed methodology, and results are compared against the ABAQUS solution times under same number of time steps and geometry, initial, loading and boundary conditions. To achieve parallel execution of the algorithm, the proposed method is implemented using OpenMP to utilise the multi-thread computing power of the chosen CPU. To be comparable, the ABAQUS solution procedure is parallelised using the in-software parallel execution settings (8 multiple processors; mode: threads). The simulation is performed at $\Delta t = 0.005\ s$ for 2000 time steps for simulation of $t = 10.0\ s$ Pennes bio-heat transfer process. By considering the TD thermal properties and parallel execution, Fig. 8 illustrates a comparison of computation times between the proposed methodology and ABAQUS utilising the six different mesh sizes. It is worth noting that the TD case (see Eq. (2)) will consume more computation time than the TI case (see Eq. (1)) since the TD thermal properties need to be updated at each time step due to the variations of tissue temperature whereas the TI thermal properties can be treated as constants and hence do not require the update. It can be seen from Fig. 8 that the proposed methodology is able to perform a calculation at $t = 3.022\ ms$ per time step at (7872 nodes, 40021 elements) and complete the simulation (2000 time steps) in $6044\ ms$ for the TD case under the OpenMP parallel execution. However, the ABAQUS takes $t = 325.50\ ms$ per time step and completes the simulation in $6.51 \times 10^5\ ms$, resulting in a computation time reduction of 107.71 times using the proposed methodology over the ABAQUS solution method for the TD case under parallelisation, far exceeding the computational performance of the commercial FEM codes. Furthermore, if no parallelisation is considered for the ABAQUS solution procedure, it takes $t = 829.75\ ms$ per time step and completes the simulation in $1.6595 \times 10^6\ ms$, leading to a computation time reduction of 274.57 times using the proposed parallel methodology over the ABAQUS non-parallel solution method for the TD case. If only TI thermal properties are considered, it takes merely $t = 1.074\ ms$ per time step at (7872 nodes, 40021 elements) and completes the simulation in $2148\ ms$ under the OpenMP parallel execution, leading to a computation time reduction of 2.81 times compared to the paralleled TD case, 303.07 times compared to the paralleled ABAQUS solution, and 772.58 times compared to the non-paralleled ABAQUS solution. A summary of computational time reductions is provided in Table. 6. It is also noticed in Fig. 8 that the computation time of the proposed methodology is increased near linearly with the increase of the system's degrees of freedom. With the physical time step size $\Delta t = 0.005\ s = 5\ ms$, the proposed methodology can achieve real-time computational performance at (7872 nodes, 40021 tetrahedrons, $t = 3.022\ ms$ per step) for the TD case with the mentioned hardware. Using a linear interpolation, it is estimated that (12823 nodes, 65991 tetrahedrons) can be simulated in real-time at $t = 5\ ms$ per step using the proposed methodology for the TD case under parallelisation.

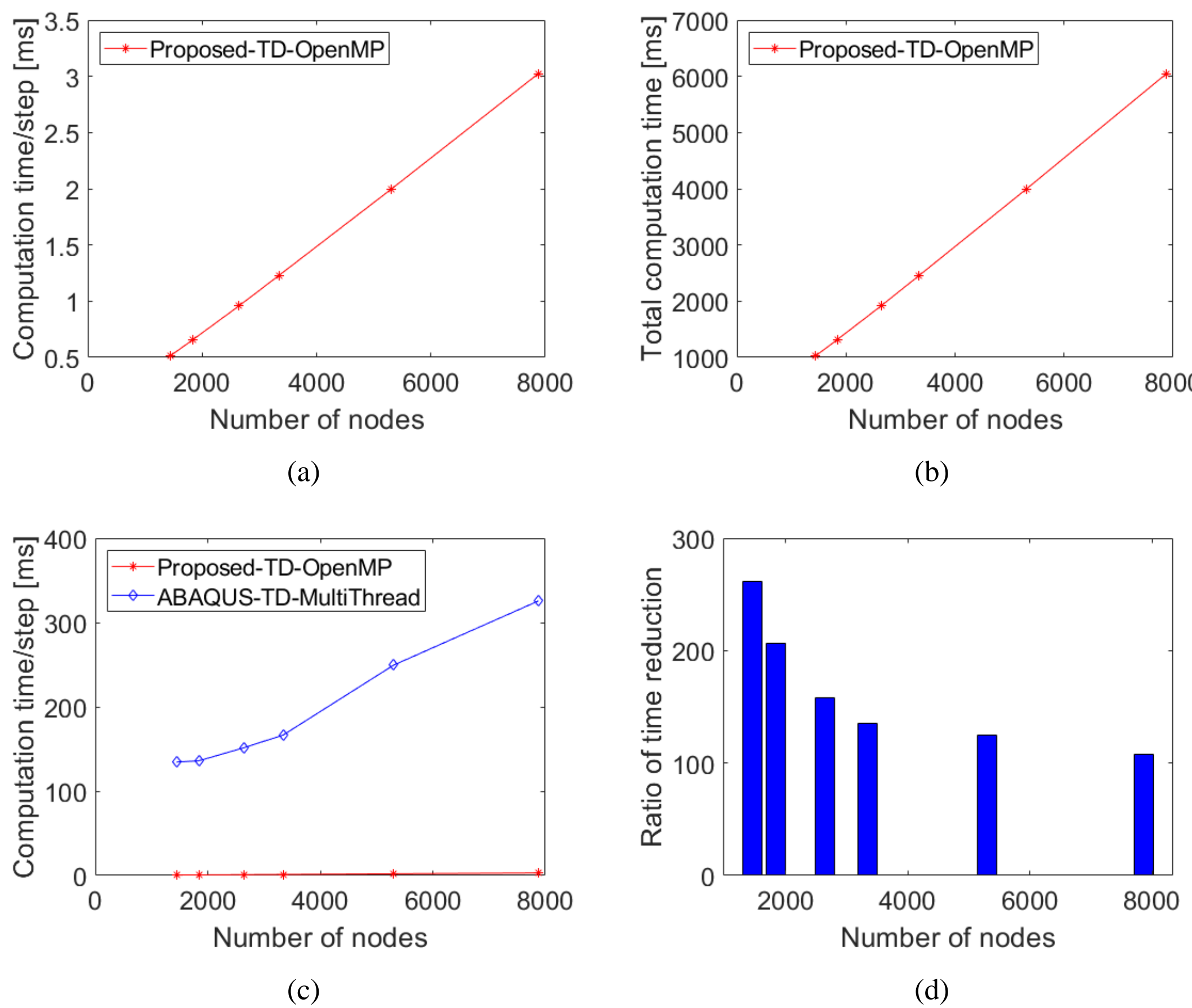

Fig. 8. Computation time of the proposed methodology and comparisons to the ABAQUS solution times utilising different mesh sizes for temperature-dependent (TD) thermal properties under parallelization (1 degree of freedom per node): (a) computation time per time step; (b) total computation time of the proposed method; (c) comparison of computation time per time step, and (d) ratio of computation time reduction ($t_{ABAQUS-Multithread}/t_{Proposed-OpenMP}$).

Table. 6

A summary of computational time reductions: P (Proposed), and A (ABAQUS).

| | $P_{TD,OpenMP}$ | $A_{threads}$ | $A_{serial}$ |
|---|---|---|---|
| $P_{TI,OpenMP}$ | 2.81x | 303.07x | 772.58x |
| $P_{TD,OpenMP}$ | N/A | 107.71x | 274.57x |

Fig. 9 illustrates a comparison of computation times of the proposed methodology with and without OpenMP parallel implementation utilising the six mesh sizes for the TD case. At (7872 nodes, 40021 elements), the proposed methodology with and without CPU parallel implementation consumes total simulation times $t = 6044\ ms$ and $t = 10002\ ms$, respectively, with a computation time $t = 3.022\ ms$ and $t = 5.001\ ms$ per time step. The CPU parallel implementation of the proposed method enables a computation time reduction of 1.655 times at (7872 nodes, 40021 elements) for the TD case. The same implementation enables a computation time reduction of 2.322 times for the TI case. The computational performance may be further improved by GPU parallel implementation [30].

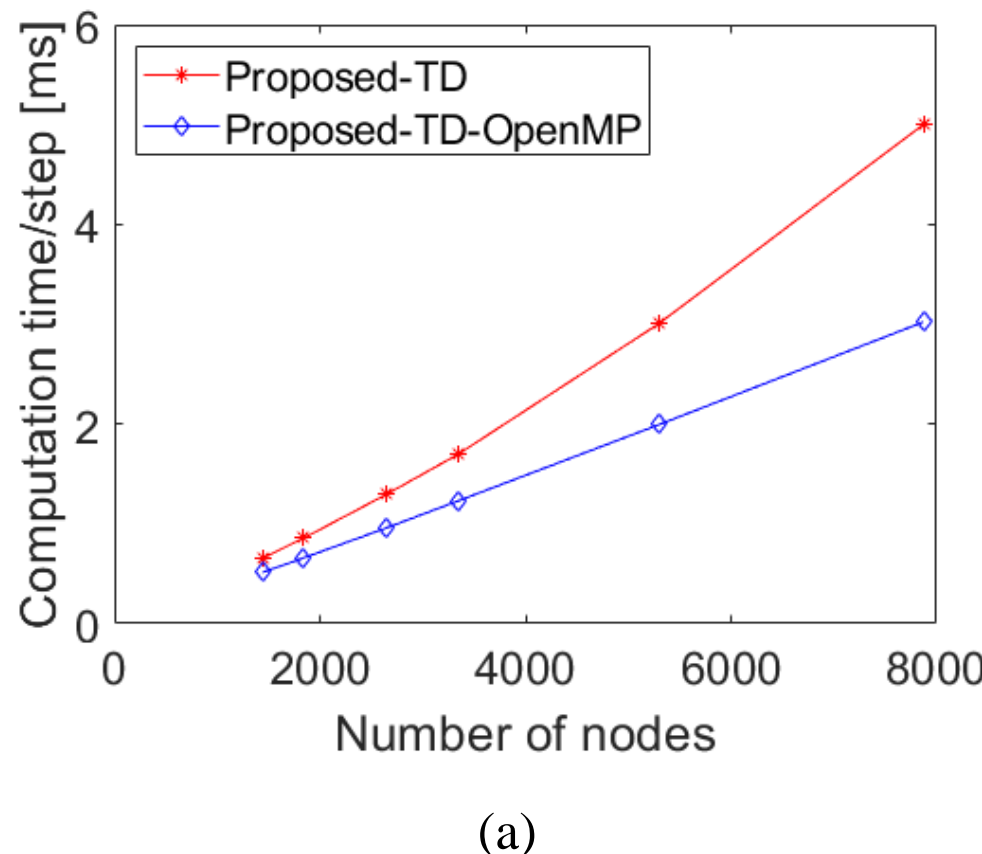

(a)

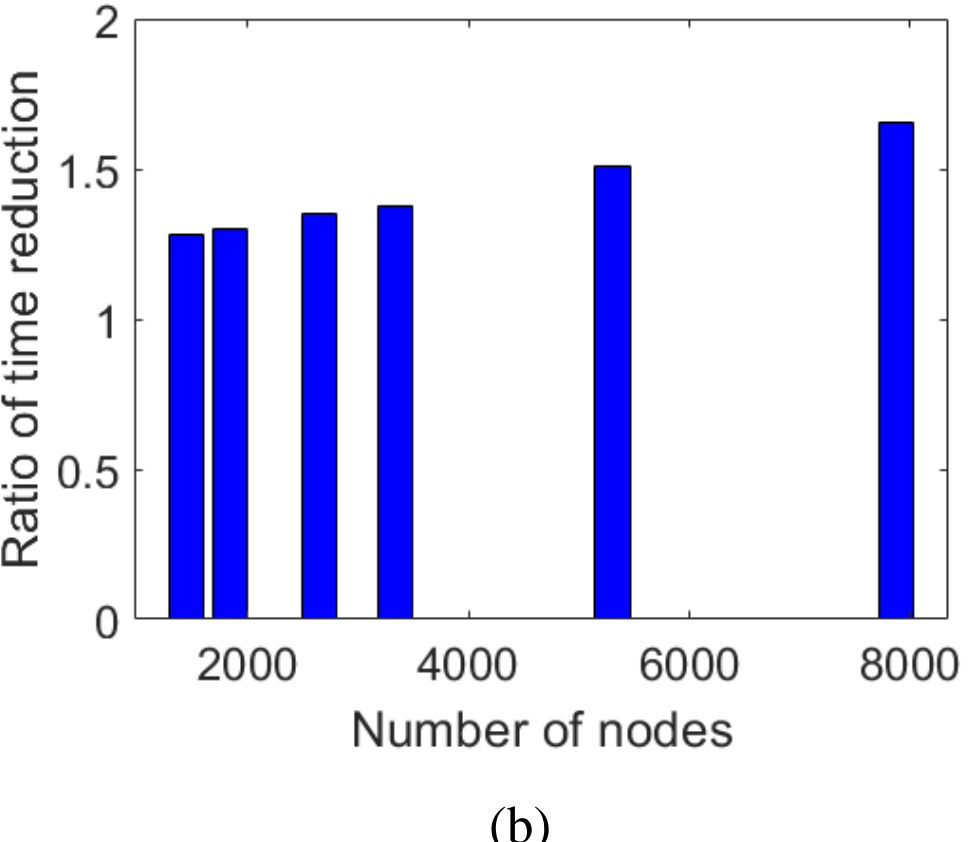

(b)

Fig. 9. Comparison of computation times between the proposed methodology with and without CPU parallel implementation utilising different mesh sizes for the TD case: (a) computation time per time step, and (b) ratio of computation time reduction ($t_{Proposed}/t_{Proposed-OpenMP}$).

## 4. Discussion

As mentioned previously, it was reported that simulation of a thermal ablation process using conventional solution methods was a time-consuming task, taking computation times ranging from a few minutes to several hours [14]. Despite the improved computational performance by the existing techniques, most of which are dominated by FDM-based techniques that suffer from inaccuracy in describing the thermal effects of irregular boundary conditions due to using FDM for computation grid. In contrast, the proposed methodology is based on the FEM discretisation that can directly handle irregular spatial domains and enforce boundary conditions. Most importantly, it provides an effective means for fast computation of bio-heat transfer in soft biological tissue and is particularly suited for real-time applications. The proposed methodology achieves fast computation of the Pennes bio-heat transfer equation via the (i) explicit dynamics in temporal domain to eliminate the need for inversion of thermal stiffness matrix for time stepping, (ii) element-level thermal load computation to eliminate the need for assembling thermal stiffness matrix of the entire model, (iii) computationally efficient finite elements to efficiently obtain the thermal responses in the spatial domain, (iv) explicit formulation for unknown nodal temperature to eliminate the need for conventional iterative process anywhere in the algorithm and permit numerical calculation to be solved independently and implemented in parallel, and (v) pre-computation of simulation parameters to further promote the computational efficiency of run-time computation.

Using the OpenMP parallel implementation of the proposed methodology on the mentioned hardware and considering TD thermal properties, simulation results reveal that the proposed method can achieve computation time reduction of 107.71 and 274.57 times compared to those of with and without parallelisation of the commercial finite element codes while maintaining a typical 1.0e-4 level of numerical accuracy. If TI thermal properties are considered under the parallel execution, it can achieve computation time reduction of 2.81, 303.07 and 772.58 times compared to the paralleled TD execution, with and without parallelisation of the commercial FEM codes, respectively, far exceeding the computational performance of the commercial codes. With the simulation time step size $\Delta t = 0.005\ s = 5\ ms$, the real-time computational performance for the TD case is achieved at (7872 nodes, 40021 tetrahedrons, $t = 3.022\ ms$ per step), and it is estimated to simulate (12823 nodes, 65991 tetrahedrons) in real-time at $t = 5\ ms$ per step using a linear interpolation; thus, the proposed algorithm constitutes a step towards real-time bio-heat transfer applications. The results from these finite element experiments are promising, demonstrating not only a high level of precision but also a significant reduction in computation time can be achieved using the proposed methodology, and it is well suited for performing real-time simulation. Some newly developed numerical methods such as the 3D alpha FEM [12] and edge-based smoothed FEM [13] are focused more on the numerical accuracy, convergence, and stability rather than the real-time computational performance, the proposed algorithm may be incorporated into these methods to enable real-time Pennes bio-heat transfer simulation. In terms of numerical stability, it was reported that neural networks [53, 54] can achieve stable simulation in the dynamic mechanical systems; they may be incorporated into the proposed method for stable dynamic bio-heat transfer.

## 5. Conclusion

This paper presents a fast numerical algorithm based on FED-FEM framework for analysis of Pennes bio-heat transfer in soft tissue, and it is particularly suited for real-time applications. The proposed methodology can achieve fast computation of Pennes bio-heat transfer owing to (i) formulating the temporal dynamics of the Pennes bio-heat transfer equation based on the explicit dynamics for efficient temporal integration, (ii) devising an element-level computation for nodal thermal loads, eliminating the need for assembling the global thermal matrices, and further (iii) utilising computationally efficient finite elements for efficient computation of thermal responses in the spatial domain, subsequently leading to (iv) an explicit formulation for nodal temperature calculation to eliminate the need for conventional iterative solution process anywhere in the algorithm and allow parallel implementation, and (v) pre-computing the constant matrices and simulation parameters to facilitate run-time computational performance. Nonlinear thermal material properties and nonlinear thermal boundary conditions can be accommodated. Simulations and comparative analyses demonstrate that the proposed method can achieve significant computation time reductions compared to those of the commercial finite element codes while maintaining good agreement in numerical accuracy, capable of achieving high computational performance for real-time simulation and analysis of Pennes bio-heat transfer problems. The clinical application of the proposed methodology could be demonstrated using various modalities for thermal treatment. Thermal therapy uses energy sources, such as radiofrequency, microwave, laser, and focused ultrasound, to generate heat to cause thermal injury to selective targets such as tumours/cancers. These thermal therapies require knowledge of tissue temperature for cancer treatment using sufficient thermal energy while not damaging surrounding healthy tissue due to overheating.

Future research work will focus on improvement and extension of the proposed methodology in two aspects. Firstly, the proposed methodology will be extended for integration with GPU hardware to further facilitate computational performance. The current implementation only utilises the computing power of CPU via the parallel implementation of OpenMP. It is expected that the GPU-accelerated solution method will further improve the computational performance owing to the massively parallel architecture of GPU consisting of thousands of cores that handle tasks simultaneously. Secondly, the proposed methodology will be extended for treatment planning and thermal dose computation using a specific modality and its associated propagation equations in tissue to achieve real-time thermal dose optimisation for better clinical outcomes of hyperthermia and thermal ablative treatments.

## Acknowledgment

This work is funded by the Australian National Health and Medical Research Council (NHMRC) Grant APP1093314.

## References

[1] H. P. Kok, A. Kotte, and J. Crezee, "Planning, optimisation and evaluation of hyperthermia treatments," *Int J Hyperthermia,* vol. 33, no. 6, pp. 593-607, 2017.

[2] B. R. Loiola, H. R. Orlande, and G. S. Dulikravich, “Thermal damage during ablation of biological tissues,” *Numerical Heat Transfer, Part A: Applications,* vol. 73, no. 10, pp. 685-701, 2018.

[3] N. Afrin, and Y. Zhang, “Uncertainty analysis of thermal damage to living biological tissues by laser irradiation based on a generalized duel-phase lag model,” *Numerical Heat Transfer, Part A: Applications,* vol. 71, no. 7, pp. 693-706, 2017.

[4] J. Zhang, Y. Zhong, and C. Gu, “Deformable Models for Surgical Simulation: A Survey,” *IEEE Rev Biomed Eng,* vol. 11, pp. 143-164, 2018.

[5] M. M. Paulides, P. R. Stauffer, E. Neufeld, P. F. Maccarini, A. Kyriakou, R. A. Canters, C. J. Diederich, J. F. Bakker, and G. C. Van Rhoon, “Simulation techniques in hyperthermia treatment planning,” *Int J Hyperthermia,* vol. 29, no. 4, pp. 346-57, 2013.

[6] V. Lopresto, R. Pinto, L. Farina, and M. Cavagnaro, “Treatment planning in microwave thermal ablation: clinical gaps and recent research advances,” *Int J Hyperthermia,* vol. 33, no. 1, pp. 83-100, 2017.

[7] H. P. Kok, P. Wust, P. R. Stauffer, F. Bardati, G. C. van Rhoon, and J. Crezee, “Current state of the art of regional hyperthermia treatment planning: a review,” *Radiat Oncol,* vol. 10, no. 1, pp. 196, 2015.

[8] S. C. Wu, G. R. Liu, H. O. Zhang, X. Xu, and Z. R. Li, “A node-based smoothed point interpolation method (NS-PIM) for three-dimensional heat transfer problems,” *International Journal of Thermal Sciences,* vol. 48, no. 7, pp. 1367-1376, 2009.

[9] K. Yang, G.-H. Jiang, H.-Y. Li, Z.-b. Zhang, and X.-W. Gao, “Element differential method for solving transient heat conduction problems,” *International Journal of Heat and Mass Transfer,* vol. 127, pp. 1189-1197, 2018.

[10] Z. C. Li, X. Y. Cui, and Y. Cai, “Analysis of heat transfer problems using a novel low-order FEM based on gradient weighted operation,” *International Journal of Thermal Sciences,* vol. 132, pp. 52-64, 2018.

[11] T. J. Yang, and X. Y. Cui, “A random field model based on nodal integration domain for stochastic analysis of heat transfer problems,” *International Journal of Thermal Sciences,* vol. 122, pp. 231-247, 2017.

[12] E. Li, G. R. Liu, V. Tan, and Z. C. He, “Modeling and simulation of bioheat transfer in the human eye using the 3D alpha finite element method (αFEM),” *International Journal for Numerical Methods in Biomedical Engineering,* vol. 26, no. 8, pp. 955-976, 2010.

[13] E. Li, G. R. Liu, and V. Tan, “Simulation of Hyperthermia Treatment Using the Edge-Based Smoothed Finite-Element Method,” *Numerical Heat Transfer, Part A: Applications,* vol. 57, no. 11, pp. 822-847, 2010.

[14] C. Rieder, T. Kroger, C. Schumann, and H. K. Hahn, “GPU-based real-time approximation of the ablation zone for radiofrequency ablation,” *IEEE Trans Vis Comput Graph,* vol. 17, no. 12, pp. 1812-21, 2011.

[15] T. Kröger, I. Altrogge, T. Preusser, P. L. Pereira, D. Schmidt, A. Weihusen, and H.-O. Peitgen, "Numerical Simulation of Radio Frequency Ablation with State Dependent Material Parameters in Three Space Dimensions," *Medical Image Computing and Computer-Assisted Intervention – MICCAI 2006.* pp. 380-388.

[16] C. W. Huang, M. K. Sun, B. T. Chen, J. Shieh, C. S. Chen, and W. S. Chen, “Simulation of thermal ablation by high-intensity focused ultrasound with temperature-dependent properties,” *Ultrason Sonochem,* vol. 27, pp. 456-465, 2015.

[17] S. R. Guntur, K. I. Lee, D. G. Paeng, A. J. Coleman, and M. J. Choi, “Temperature-dependent thermal properties of ex vivo liver undergoing thermal ablation,” *Ultrasound Med Biol,* vol. 39, no. 10, pp. 1771-84, 2013.

[18] J. Zhang, J. Hills, Y. Zhong, B. Shirinzadeh, J. Smith, and C. Gu, “Temperature-Dependent Thermomechanical Modeling of Soft Tissue Deformation,” *Journal of Mechanics in Medicine and Biology,* vol. 18, no. 08, pp. 1840021, 2019.

[19] T. Wegner, and A. Pęczak, “Implementation of a strain energy-based nonlinear finite element in the object-oriented environment,” *Computer Physics Communications,* vol. 181, no. 3, pp. 520-531, 2010.

[20] M. Schwenke, J. Georgii, and T. Preusser, “Fast Numerical Simulation of Focused Ultrasound Treatments During Respiratory Motion With Discontinuous Motion Boundaries,” *IEEE Trans Biomed Eng,* vol. 64, no. 7, pp. 1455-1468, 2017.

[21]Z.-Z. He, and J. Liu, “An efficient parallel numerical modeling of bioheat transfer in realistic tissue structure,” *International Journal of Heat and Mass Transfer,* vol. 95, pp. 843-852, 2016.

[22]G. Carluccio, D. Erricolo, S. Oh, and C. M. Collins, “An approach to rapid calculation of temperature change in tissue using spatial filters to approximate effects of thermal conduction,” *IEEE Trans Biomed Eng,* vol. 60, no. 6, pp. 1735-41, 2013.

[23]G. Kalantzis, W. Miller, W. Tichy, and S. LeBlang, "A GPU accelerated finite differences method of the bioheat transfer equation for ultrasound thermal ablation," *Software Engineering, Artificial Intelligence, Networking and Parallel/Distributed Computing*, pp. 45-55: Springer, 2016.

[24]J. L. Dillenseger, and S. Esneault, “Fast FFT-based bioheat transfer equation computation,” *Comput Biol Med,* vol. 40, no. 2, pp. 119-23, 2010.

[25]G. Chen, J. Stang, M. Haynes, E. Leuthardt, and M. Moghaddam, “Real-Time Three-Dimensional Microwave Monitoring of Interstitial Thermal Therapy,” *IEEE Trans Biomed Eng,* vol. 65, no. 3, pp. 528-538, 2018.

[26]P. C. Johnson, and G. M. Saidel, “Thermal model for fast simulation during magnetic resonance imaging guidance of radio frequency tumor ablation,” *Ann Biomed Eng,* vol. 30, no. 9, pp. 1152-61, 2002.

[27]J. H. Niu, H. Z. Wang, H. X. Zhang, J. Y. Yan, and Y. S. Zhu, “Cellular neural network analysis for two-dimensional bioheat transfer equation,” *Med Biol Eng Comput,* vol. 39, no. 5, pp. 601-4, 2001.

[28]J. H. Indik, R. A. Indik, and T. C. Cetas, “Fast and efficient computer modeling of ferromagnetic seed arrays of arbitrary orientation for hyperthermia treatment planning,” *Int J Radiat Oncol Biol Phys,* vol. 30, no. 3, pp. 653-62, 1994.

[29]R. A. Indik, and J. H. Indik, “A new computer method to quickly and accurately compute steady-state temperatures from ferromagnetic seed heating,” *Med Phys,* vol. 21, no. 7, pp. 1135-44, 1994.

[30]P. Mariappan, P. Weir, R. Flanagan, P. Voglreiter, T. Alhonnoro, M. Pollari, M. Moche, H. Busse, J. Futterer, H. R. Portugaller, R. B. Sequeiros, and M. Kolesnik, “GPU-based RFA simulation for minimally invasive cancer treatment of liver tumours,” *Int J Comput Assist Radiol Surg,* vol. 12, no. 1, pp. 59-68, 2017.

[31]J. Zhang, J. Hills, Y. Zhong, B. Shirinzadeh, J. Smith, and C. Gu, “GPU-Accelerated Finite Element Modeling of Bio-Heat Conduction for Simulation of Thermal Ablation,” *Journal of Mechanics in Medicine and Biology,* vol. 18, no. 07, pp. 1840012, 2018.

[32]G. C. Bourantas, M. Ghommem, G. C. Kagadis, K. Katsanos, V. C. Loukopoulos, V. N. Burganos, and G. C. Nikiforidis, “Real-time tumor ablation simulation based on the dynamic mode decomposition method,” *Med Phys,* vol. 41, no. 5, pp. 053301, 2014.

[33]M. E. Kowalski, and J. M. Jin, “Model-order reduction of nonlinear models of electromagnetic phased-array hyperthermia,” *IEEE Trans Biomed Eng,* vol. 50, no. 11, pp. 1243-54, 2003.

[34]C. Ding, X. Cui, R. R. Deokar, G. Li, Y. Cai, and K. K. Tamma, “An isogeometric independent coefficients (IGA-IC) reduced order method for accurate and efficient transient nonlinear heat conduction analysis,” *Numerical Heat Transfer, Part A: Applications,* vol. 73, no. 10, pp. 667-684, 2018.

[35]J. Zhang, Y. Zhong, and C. Gu, “Energy balance method for modelling of soft tissue deformation,” *Computer-Aided Design,* vol. 93, pp. 15-25, 2017.

[36]J. Zhang, and S. Chauhan, “Fast explicit dynamics finite element algorithm for transient heat transfer,” *International Journal of Thermal Sciences,* vol. 139, pp. 160-175, 2019.

[37]H. H. Pennes, “Analysis of tissue and arterial blood temperatures in the resting human forearm,” *J Appl Physiol,* vol. 1, no. 2, pp. 93-122, 1948.

[38]W. Shen, J. Zhang, and F. Yang, “Modeling and numerical simulation of bioheat transfer and biomechanics in soft tissue,” *Mathematical and Computer Modelling,* vol. 41, no. 11-12, pp. 1251-1265, 2005.

[39]J. Durkee Jr, and P. Antich, “Exact solutions to the multi-region time-dependent bioheat equation with transient heat sources and boundary conditions,” *Physics in medicine & biology,* vol. 36, no. 3, pp. 345, 1991.

[40]P. Prakash, and C. J. Diederich, "Considerations for theoretical modelling of thermal ablation with catheter-based ultrasonic sources: implications for treatment planning, monitoring and control," *Int J Hyperthermia,* vol. 28, no. 1, pp. 69-86, 2012.

[41]Z.-Z. He, and J. Liu, "A coupled continuum-discrete bioheat transfer model for vascularized tissue," *International Journal of Heat and Mass Transfer,* vol. 107, pp. 544-556, 2017.

[42]A. R. A. Khaled, and K. Vafai, "The role of porous media in modeling flow and heat transfer in biological tissues," *International Journal of Heat and Mass Transfer,* vol. 46, no. 26, pp. 4989-5003, 2003.

[43]E. L. Wilson, K. J. Bathe, and F. E. Peterson, "Finite element analysis of linear and nonlinear heat transfer," *Nuclear Engineering and Design,* vol. 29, no. 1, pp. 110-124, 1974.

[44]X. Rong, R. Niu, and G. Liu, "Stability Analysis of Smoothed Finite Element Methods with Explicit Method for Transient Heat Transfer Problems," *International Journal of Computational Methods*, pp. 1845005, 2018.

[45]E. Li, Z. C. He, Q. Tang, and G. Y. Zhang, "Large time steps in the explicit formulation of transient heat transfer," *International Journal of Heat and Mass Transfer,* vol. 108, pp. 2040-2052, 2017.

[46]X.-G. Li, L.-P. Yi, Z.-Z. Yang, Y.-T. Chen, and J. Sun, "Coupling model for calculation of transient temperature and pressure during coiled tubing drilling with supercritical carbon dioxide," *International Journal of Heat and Mass Transfer,* vol. 125, pp. 400-412, 2018.

[47]R. Courant, K. Friedrichs, and H. Lewy, "On the partial difference equations of mathematical physics," *IBM journal of Research and Development,* vol. 11, no. 2, pp. 215-234, 1967.

[48]K.-J. Bathe, *Finite element procedures*: Klaus-Jurgen Bathe, 2006.

[49]E. Li, Z. C. He, Z. Zhang, G. R. Liu, and Q. Li, "Stability analysis of generalized mass formulation in dynamic heat transfer," *Numerical Heat Transfer, Part B: Fundamentals,* vol. 69, no. 4, pp. 287-311, 2016.

[50]S. Haddadi, and M. T. Ahmadian, "Numerical and Experimental Evaluation of High-Intensity Focused Ultrasound-Induced Lesions in Liver Tissue Ex Vivo," *J Ultrasound Med,* vol. 37, no. 6, pp. 1481-1491, 2018.

[51]P. Rattanadecho, and P. Keangin, "Numerical study of heat transfer and blood flow in two-layered porous liver tissue during microwave ablation process using single and double slot antenna," *International Journal of Heat and Mass Transfer,* vol. 58, no. 1-2, pp. 457-470, 2013.

[52]W. Karaki, Rahul, C. A. Lopez, D. A. Borca-Tasciuc, and S. De, "A continuum thermomechanical model of in vivo electrosurgical heating of hydrated soft biological tissues," *Int J Heat Mass Transf,* vol. 127, no. Pt A, pp. 961-974, 2018.

[53]J. Zhang, Y. Zhong, J. Smith, and C. Gu, "Cellular neural network modelling of soft tissue dynamics for surgical simulation," *Technol Health Care,* vol. 25, no. S1, pp. 337-344, 2017.

[54]J. Zhang, Y. Zhong, and C. Gu, "Neural network modelling of soft tissue deformation for surgical simulation," *Artif Intell Med*, 2018.